\input ./style/arxiv-general.cfg
\documentclass[aoas,MSNbibl,nameyear,dvips]{arximspdf}
\makeatletter
   \@ifpackageloaded{graphicx}{}{\usepackage{graphicx}}
\makeatother
\usepackage{dcolumn}

\doi{10.1214/15-AOAS845}% Updated by VTEXPTS2LaTeX.exe, 14.07.2015 15:49
\volume{9}
\issue{3}
\pubyear{2015}
\firstpage{1415}
\lastpage{1432}
\docsubty{FLA}

\makeatletter
\newcolumntype{d}[1]{D{.}{.}{#1}}
\def\mid{|}
\makeatother

\begin{document}
\begin{frontmatter}

\title{A Bayesian approach to the evaluation of risk-based
microbiological criteria for \emph{\uppercase{Campylobacter}} in broiler meat}
\runtitle{risk-based criteria for \emph{\uppercase{Campylobacter}}}

\begin{aug}
% Corresponding author: Jukka Ranta - jukka.ranta@evira.fi% Updated by
%VTEXPTS2LaTeX.exe, 14.07.2015 15:49
\author[A]{\fnms{Jukka} \snm{Ranta}\corref{}\ead[label=e1]{jukka.ranta@evira.fi}\thanksref{t1}},
\author[B]{\fnms{Roland} \snm{Lindqvist}\ead[label=e2]{roland.lindqvist@slv.se}\thanksref{t2,t22}},
\author[C]{\fnms{Ingrid} \snm{Hansson}\ead[label=e3]{ingrid.hansson@sva.se}\thanksref{t3}},\\
\author[D]{\fnms{Pirkko} \snm{Tuominen}\ead[label=e4]{pirkko.tuominen@evira.fi}\thanksref{t1}}
\and
\author[E]{\fnms{Maarten} \snm{Nauta}\ead[label=e5]{maana@food.dtu.dk}\thanksref{t4}}
\runauthor{J. Ranta et al.}
\affiliation{Finnish Food Safety Authority Evira\thanksmark{t1},
National Food Agency\thanksmark{t2},
Swedish University of Agricultural Sciences\thanksmark{t22}
National Veterinary Institute\thanksmark{t3} and
Technical University of Denmark\thanksmark{t4}}
\address[A]{J. Ranta\\
Risk Assessment Research Unit \\
Finnish Food Safety Authority Evira\\
Mustialankatu 3 \\
FIN-00790 Helsinki\\
Finland \\
\printead{e1}}
\address[B]{R. Lindqvist \\
National Food Agency \\
P.O. Box 622 \\
SE-75126 Uppsala\\
Sweden\\
and\\
Swedish University of Agricultural Sciences SLU\\
P.O. Box 7070 \\
SE-75007 Uppsala\\
Sweden \\
\printead{e2}}
\address[C]{I. Hansson\\
National Veterinary Institute SVA\\
SE-75189 Uppsala\\
Sweden\\
\printead{e3}}
\address[D]{P. Tuominen \\
Risk Assessment Research Unit \\
Finnish Food Safety Authority Evira\hspace*{52pt}\\
Mustialankatu 3 \\
FIN-00790 Helsinki\\
Finland \\
\printead{e4}}
\address[E]{M. Nauta \\
National Food Institute \\
Technical University of Denmark DTU \\
M{\o}rkh{\o}j Bygade 19 \\
DK-2860 S{\o}borg\\
Denmark \\
\printead{e5}}

%\runauthor{}
%%\dedicated{}
\end{aug}

% HISTORY:
%
\received{\smonth{3} \syear{2014}}% Updated by VTEXPTS2LaTeX.exe,
%14.07.2015 15:49
%
\revised{\smonth{5} \syear{2015}}% Updated by VTEXPTS2LaTeX.exe,
%14.07.2015 15:49

% ABSTRACT
%
\begin{abstract}
Shifting from traditional hazard-based food safety management toward
risk-based management requires statistical methods for evaluating
intermediate targets in food production, such as microbiological
criteria (MC), in terms of their effects on human risk of illness. A
fully risk-based evaluation of MC involves several uncertainties that
are related to both the underlying Quantitative Microbiological Risk
Assessment (QMRA) model and the production-specific sample data on the
prevalence and concentrations of microbes in production batches.
We used Bayesian modeling for statistical inference and evidence
synthesis of two sample data sets. Thus, parameter uncertainty was
represented by a joint posterior distribution, which we then used to
predict the risk and to evaluate the criteria for acceptance of
production batches. We also applied the Bayesian model to compare
alternative criteria, accounting for the statistical uncertainty of
parameters, conditional on the data sets. Comparison of the posterior
mean relative risk, $E(\mathit{RR} \mid\mathrm{data}) = E(P(\mathrm{illness} \mid
\mathrm{criterion\ is\ met})/P(\mathrm{illness}) \mid\mathrm{data})$, and
relative posterior risk, $\mathit{RPR}=P(\mathrm{illness} \mid\mathrm{data,\
criterion\ is\ met})/P(\mathrm{illness} \mid\mathrm{data})$, showed very
similar results, but computing is more efficient for RPR. Based on the
sample data, together with the QMRA model, one could achieve a relative
risk of 0.4 by insisting that the default criterion be fulfilled for
acceptance of each batch.
\end{abstract}

% KEYWORDS
\end{frontmatter}

%s1 #&#
\section{Introduction}

Campylobacteriosis is the most commonly reported bacterial enteric
disease in humans in many industrial countries [\citet{efsa2013}]. One
risk factor for human campylobacteriosis is handling and consuming
contaminated poultry meat [\citet{kapperud}; \citet{wingstrand}].
During slaughter, broiler carcasses can become contaminated with
\textit{Campylobacter}, and this contamination in the slaughter batch can
originate from the intestinal contents or from the environment [\citeauthor{rosenquist2} (\citeyear{rosenquist2,rosenquist});
\citet{lindqvist2008}; \citet
{nauta2009}].

Quantitative Microbiological Risk Assessments (QMRA) of food-borne
pathogens in a production chain aim to provide numerical estimates of
consumer risk. They involve modeling several steps in the production
chain or a complete farm-to-fork chain starting from the primary
production models and ending with dose-response models [\citet
{nauta2009}]. At best, the available data can only partially cover the steps.
A common approach uses detailed simulation models that aim to provide a
realistic representation of the mechanistic nature of various processes
known to influence the survival of pathogens and their transmission,
leading to possible consumer exposure. Other, parsimonious statistical
models aim to bridge the data gaps by using the observed data to
provide estimates of the overall effects without making too many assumptions.
Detailed models inevitably involve many parameters. Because of limited
data, however, the resulting estimates can have considerable
uncertainty, which is often suppressed, and can lead to predictions
whose uncertainties are more assumption-driven than data-driven.
Although assumptions can provide insight for possible or hypothetical
what-if scenarios, data can be more fully exploited through formal
statistical inference, which in turn can provide estimates with
uncertainty bounds based on the empirical evidence.

Since the Bayesian interpretation of probability is necessarily
conditional and depends on the available evidence (prior${} + {}$empirical
data), it provides a logical way to assess multidimensional parameter
uncertainty that can be explicitly updated by new data [\citet
{gelmanhill}; \citet{SpiegelhalterBest}], specifically in the context
of microbiological risk assessments [\citet{naujas}] addressing bacterial growth [\citet
{spor}] or management interventions for better food safety [\citet
{ranta2002}; \citeauthor{ranta2010} (\citeyear{ranta2010,ranta2013})]. Even so, some parameters may
be so inherent in the problem that they need to be included regardless
of whether sufficient, or any, data exist. Hence, they become part of
the model uncertainty, which can be considered one of the several
levels of uncertainty [\citet{Spiegelhalter}].

We present a Bayesian method for evaluating and comparing the effects
of microbiological criteria (MC) in broiler production on consumer
risk. Microbiological criteria have been recognized as practical
measures for defining the level of acceptability in food product
testing for decades, with the earliest versions already available in
the 1960s [\citet{codex}; \citet{uscouncil}]. Our evaluation of MC is
based on uncertainty analysis concerning model parameters, for which
national sample data are available on broiler carcasses. We analyze two
types of such data combined in tandem with the subsequent QMRA model of
consumer risk. The computations were implemented using freely available
OpenBUGS software in the R environment [\citet{bugsbook}],
(\url{http://www.openbugs.net/}), and the model code is available in
the supplementary material (Section A.5) [\citet{ranta2015}].\vadjust{\goodbreak}

%f1 #&#
\begin{figure}

\includegraphics{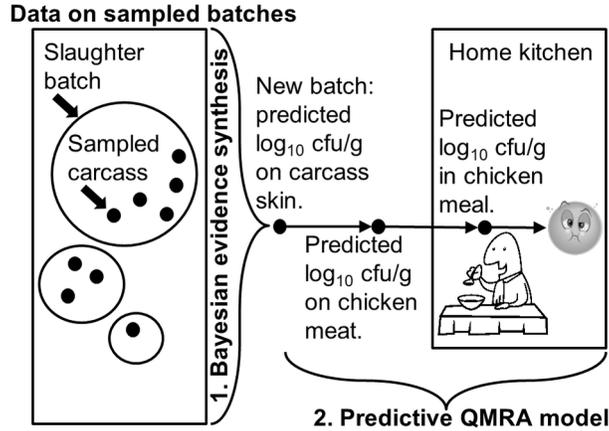}

\caption{Simplified diagram of the food chain from slaughter batches to
consumption, covering the steps
modeled. Data from slaughter batches are used for Bayesian evidence
synthesis (1). The resulting posterior distribution of underlying
parameters is then used as an input distribution in the existing QMRA
model (2) for predicting the number of colony forming units of bacteria
per gram (cfu/g) of meat and the consequent risk of illness. Another predictive
QMRA model could be used instead of the one adopted here.} \label{schema}
\end{figure}

%s2 #&#
\section{Methods and data}
%s2.1 #&#
\subsection{General model structure}

Figure~\ref{schema} shows the simplified food chain from slaughter
batches to consumption, covering the main steps modeled. In the first
part, we conduct Bayesian evidence synthesis of two data sets on
sampled broiler carcasses after slaughter. Batch structures and
microbial levels have also been studied using Bayesian models in \citet
{commeau}. In the sample data sets here, a batch originates from one
broiler flock. Generally, batches in the model could represent a single
broiler flock, part of a flock or a combination of flocks.
Usually a batch contains more than 10,000 carcasses.
If contamination occurs, prevalence within a batch is usually high. For
contaminated carcasses, the concentration level varies from carcass to
carcass, and the mean concentration varies between batches. Therefore,
the hierarchical model involves parameters for both prevalence and
concentration. Generally, we use $q$ for prevalence of contaminated
batches and $p_{j}$ for within-batch prevalence in batch $j$. The
concentration model is defined for contaminated carcasses only, so the
average concentration (as $\log_{10}$) of all contaminated carcasses is
$\mu$, and the within-batch mean is $\mu_j$. Concentrations of
individual carcasses ($i=1,2,\ldots$) are modeled as N$(y_{ij} \mid\mu
_j,\sigma^{-2}_w)$, where $\sigma_w^2$ is the within-batch variance
component and $\sigma^{-2}_w=\tau_w$ is the corresponding precision
parameter. Likewise, batch means are modeled as N$(\mu_j \mid\mu,\sigma
^{-2}_b)$, where $\sigma^2_b$ is the between-batch variance component.
The data for this model would ideally consist of measurements from
several individual carcasses from a sample of batches. Such data are
not always available, and we show an example of Bayesian analysis with
two qualitatively very different data sets. The underlying parameters
are needed for predicting contamination of \emph{a random new carcass
from a random new batch}. A~log-normal distribution is commonly used
for microbial concentrations, and Q--Q-plots were used for checking
approximate normality. Q--Q-plots of equally many random points were
generated from the standard normal distribution to see that the
Q--Q-plot of data falls reasonably within sampling error.

In the second part, an existing QMRA model describes the subsequent
processing chain from carcasses to fresh meat, meal preparation,
consumption and the probability of illness. This subsequent QMRA model
contains a sequence of conditional distributions, which are taken as
given in \citet{nauta}; see the supplementary material Section A.2
[\citet{ranta2015}]. We treat this QMRA model as a template for
computing the risk, and the essential link to the first part is that
the parameters describing carcass contamination are input parameters in
the second part. These parameters are specific to each country, with
uncertainties depending on national carcass sample data.

Finally, Bayesian posterior predictive distributions for a random
serving from a random batch were used to study the effect of various
microbiological criteria. The criteria define critical levels of
contamination per batch so that batches can be rejected or accepted
based on sample results. As a default scenario, the criterion is
defined as ``$n=5, c=1, m=1000$,'' which means that at most one ($c$)
sample out of five ($n$) is allowed to have $\log_{10}\ \mathrm{cfu/g} >3$
($m>1000$ colony forming units per gram). Knowing whether a batch was
accepted provides additional evidence, which has an effect on the
posterior distribution of parameters for that batch and, consequently,
on the predictions for servings stemming from the batch. Notation is
given in Table~\ref{notas}.

%t1 #&#
\begin{table}
\caption{List of notation}
\label{notas}
\begin{tabular*}{\textwidth}{@{\extracolsep{\fill}}ll@{}}
\hline
$j'$ & batch index in Lindblad et al. ``1/batch'' data\\
$j''$ & batch index in Hansson et al. ``$N_{j''}/\mathrm{batch}^+$''
data\\
$j$ & generic batch index for prediction\\
$J'$ & \# positive carcasses in Lindblad et al. data.\\
$N'$ & \# sampled carcasses in Lindblad et al. data ($=$ \# sampled
batches)\\
$x_{j''}$ & \# positive carcasses in $j''$th batch in Hansson et al.
data \\
$N_{j''}$ & \# sampled carcasses in $j''$th batch in Hansson et al.
data\\
$q$ & prevalence of contaminated batches\\
$p_{j''}$ & within-batch prevalence in batch $j''$\\
$\alpha$ & parameter for distribution of within-batch prevalence\\
$\mu$ & mean $\log_{10}$~cfu/g of all contaminated carcasses\\
$\mu_j$ & mean $\log_{10}$~cfu/g of contaminated carcasses in batch $j$;
$j'$ or $j''$ or generic\\
$y_{ij}$ & $\log_{10}$~cfu/g of contaminated carcass $i$ in batch $j$;
$j'$ or $j''$ or generic\\
$I_j$ & true contamination status for a generic batch\\
$\sigma_b^2$ & between-batch variance of $\mu_j$'s; either $j'$ or $j''$
or generic\\
$\sigma_w^2$ & within-batch variance of $y_{ij}$'s; either $j'$ or $j''$
or generic\\
$\tau_b$ & between-batch precision $\sigma_b^{-2}$\\[1pt]
$\tau_w$ & within-batch precision $\sigma_{w}^{-2}$\\
$y_c$ & predicted $\log_{10}$~cfu/g for a contaminated carcass to be
used for a serving\\
$w$ & weight (g) of a broiler serving\\
$n_c$ & bacteria count in a raw broiler serving of weight $w$\\
$r$ & transfer probability for a bacteria cell from raw broiler meat to
salad\\
$d$ & bacteria count, the dose, in final serving\\
$P_0(\mathrm{ill} \mid d)$ & probability of illness (dose response)\\
$\theta_j$ & batch-specific parameters ($I_j,p_j,\mu_j$) in
predictions\\
$\theta_s$ & serving-specific parameters ($d,n_c,r,w,y_c$) in
predictions\\
$L$ & \# Monte Carlo draws of batch-specific parameters per MCMC
iteration\\
$M$ & \# Monte Carlo draws of serving-specific pars. per batch per MCMC
iteration\\
MC:$ n/c/m$ & microbiological criterion: \\ & $n=$ sample size, $c=$
max \# positives
exceeding $m$~cfu/g\\
\hline
\end{tabular*}
\end{table}

%s2.2 #&#
\subsection{Two types of carcass sample data}
It is usually not possible to devise a sampling plan beforehand to
serve the ideal data needs of a risk assessment model. Two Swedish data
sets represent the types of historical data that could be commonly
available. The first set is a one-year baseline study, conducted by
\citet{lindblad} between September 2002 and August 2003, that reports
data on the prevalence and levels of thermophilic \textit{Campylobacter}
species in Swedish broiler chickens.
%('spp' - \textit{ species pluralis} - for multiple species).
Batches were sampled from ten slaughterhouses that represent 99.9\% of
the yearly production, and the number of samples per slaughterhouse was
proportional to the annual production.
One chilled carcass per batch was analyzed, and \textit{Campylobacter} was
quantified by direct plating in 88 out of 617 carcasses. These data
have been used in another quantitative risk assessment by
\citet{lindqvist2008}. Bacteria concentrations as $\log_{10}$ values on
positive whole carcasses represent bacteria concentration per carcass
skin. According to \citet{nauta}, in order to transform the values to
$\log_{10}$~cfu per gram of skin, it was assumed that skin weight is
approximately 100~g, so $\log_{10}(100)$ was subtracted from each
measurement. The first data set, data1, also denoted by ``1/batch''
data, hence represents one sample for each of the $j'=1,\ldots,N'=617$
batches.

The second data set, presented by \citet{hansson}, describes a
collection of carcass samples taken between July and October 2006, from
20 batches delivered by producers with a history in the Swedish
\textit{Campylobacter} surveillance program of often delivering
\textit{Campylobacter}-positive broilers. All batches were positive. The sample
size per batch varied from 5 to 25, and the percentage of positives per
sample varied from 85\% to 100\% with a mean of 98\%. Sample means and
sample standard deviations for positive carcasses were reported per
batch. In these data, the measurements represent $\log_{10}$~cfu per ml
of rinse water when 400 ml of water was used. According to \citet
{nauta}, these were transformed into values per carcass by adding $\log
_{10}(400)$ to the original values, then subtracting $\log_{10}(100)$,
with the result that $\log_{10}(4)$ was added to each measurement.
Transformation to cfu/g is necessary for compatibility with
dose-response models. The second data set, data2, also denoted by
``$N_{j''}/\mathrm{batch}^+$'' data, hence represents $N_1,\ldots
,N_{J''}$ samples from $J''=20$ positive batches.

These two data sets provide complementary evidence. From the first
sample we obtain some information about total variance and overall
batch prevalence but nothing about within-batch prevalence. The second
data set provides information on within-batch parameters for positive
batches, but nothing on the overall batch prevalence. Therefore,
evidence synthesis is needed.

%s2.3 #&#
\subsection{Evidence synthesis from carcass sample data}

\begin{figure}

\includegraphics{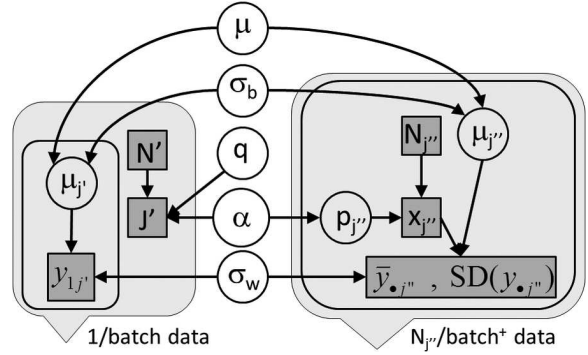}

\caption{Directed acyclic graph of the final evidence synthesis model
combining two qualitatively different data sets with common parameters
in the middle.
Incoming arrows denote conditional distributions as explained in the
text.} \label{modelDAG}
\end{figure}

A common challenge of QMRAs arises from the limited amount of data
available. Typically, only one data set of the types presented here may
be available. These limitations have a direct influence on the
uncertainties, which can be quantified and presented as a posterior
distribution of the parameters.
Below, we present results based on both of the sample data sets taken
separately and combined to illustrate this point. The full evidence
synthesis model is shown as a directed acyclic graph (DAG) in
Figure~\ref{modelDAG}.

%f2 #&#

%s2.3.1 #&#
\subsubsection{Modeling with baseline data: One sample per batch
(``1/batch'' data)}
\begin{figure}

\includegraphics{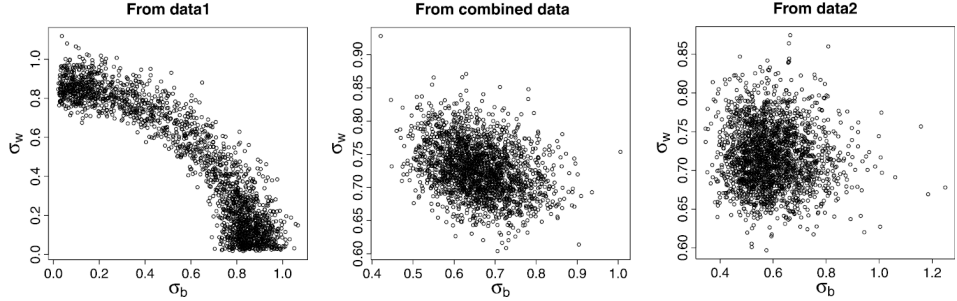}

\caption{Marginal posterior distributions of $(\sigma_b,\sigma_w)$ based
on each data set alone (1/batch left, $N_{j''}/\mathrm{batch}^+$ right)
and the two data sets combined (middle).} \label{sigfig}
\end{figure}

The first data set provides a representative baseline sample for one year.
To compute the posterior distribution of model parameters from these
data alone, one has to decide what to assume concerning within-batch
prevalence, for which there was no information at all. An expert
elicitation might provide a prior distribution, but
for the example, we use the background information that contaminated
batches have a high within-batch prevalence, and we simply set $p_{j'}
= 1$ for all such batches, $j'=1,\ldots,J'$.
From Bayes's theorem, the posterior distribution is then
%
%e1 #&#
\begin{eqnarray}&&
\pi\bigl(\mu,\tau_b,\tau_w,q,\mu_1,
\ldots,\mu_{J'} \mid J',N',y_{11},
\ldots,y_{1J'} \bigr)
\nonumber
\\
&&\qquad\propto
\operatorname{Binomial}\bigl(J' \mid N',q\bigr) \prod
_{j'=1}^{J'} \mathrm{N}(y_{1j'} \mid
\mu_{j'}, \tau_w)\\
&&\qquad\quad{}\times\mathrm{N}(\mu _{j'} \mid\mu,
\tau_b) \pi(\mu)\pi(\tau_b)\pi(\tau_w)
\pi(q),\nonumber
\end{eqnarray}
where $N'=617, J{'}=88$, and $y_{1j'}$ are the adjusted $\log_{10}$
concentrations. Conventionally, the variance parameters were replaced
with the precision parameters $\tau_b=\sigma_b^{-2}, \tau_w=\sigma
_w^{-2}$. The posterior distribution can now be computed using the
prior distributions $q \sim\mathrm{U}(0,1)$,
$\mu\sim\mathrm{N}(0,10^{-4})$, $\tau_b \sim\operatorname{Gamma}(0.001,\break 0.001)$
and $\tau_w \sim \operatorname{Gamma}(0.001,0.001)$.
The Gamma priors resemble the uninformative priors $\pi(\tau_b) \propto
\tau_b^{-1}$, $\pi(\tau_w) \propto\tau_w^{-1}$. The Gamma prior for
between-batch precision can be problematic in some situations and
should not be an automatic choice. Sensitivity to priors is discussed below.
The normal prior for $\mu$ is practically flat, to reflect the lack of
prior knowledge. The marginal posterior distribution of $(\sigma
_b,\sigma_w)$, based on these data, is shown at the left in Figure~\ref
{sigfig}. Since ``1/batch'' data provide evidence on total variance, we
can infer what combinations of values for variance parameters are more
probable than others. If these were the only available data, the
consequent uncertainty would be described by this joint distribution,
which would be further reflected in the predictive distributions of the
log-cfu concentrations.

%f3 #&#

%s2.3.2 #&#
\subsubsection{Modeling based on more than one sample per batch, for
positive batches only (``$N_{j''}/\mathrm{batch}^+$'' data)}

With only ``$N_{j''}/\mathrm{batch}^+$'' data on positive batches, there
is no information about batch prevalence $q$. However, it is possible
to estimate within-batch prevalence $p_{j''}$ for each batch, and also
to construct
a hierarchical model for $p_{j''}$ with hyperparameter $\alpha$ to
describe variation in within-batch prevalences. Knowing that prevalence
is usually high, we restrict ourselves to distributions peaked near 1
by choosing prior $p_{j''} \sim \operatorname{Beta}(\alpha,2)$. The reported
sample means and sample standard deviations of the $\log_{10}$
concentrations summarize the observed $x_{j''}$ positive carcasses per
batch, so the posterior becomes
%
%e2 #&#
\begin{eqnarray}
&&\pi\bigl(\mu,\mu_1,\ldots,\mu_{J''},\tau_b,
\tau_w,p_1,\ldots ,p_{J''},\alpha\mid \bigl
\{x_{j''},N_{j''},\bar{y}_{\cdot j''},\mathrm{SD}(y_{\cdot
j''})\bigr\}_{j''=1,\ldots,J''} \bigr)\nonumber\\
&&\qquad \propto
\prod_{j''=1}^{J''} \operatorname{Binomial}(x_{j''}
\mid N_{j''},p_{j''}) \pi\bigl( \bar{y}_{\cdot j''}, \mathrm{
SD}(y_{\cdot j''}) \mid\mu _{j''},\tau_w,x_{j''}
\bigr) \mathrm{N}(\mu_{j''} \mid\mu,\tau_b)\\
&&\qquad\quad{} \times
\operatorname{Beta}(p_{j''} \mid\alpha,2) \pi(\mu)\pi(\tau_b)\pi(
\tau_w)\pi (\alpha).\nonumber
\end{eqnarray}

Implementing the posterior is less straightforward because of the term
$\pi( \bar{y}_{\cdot j''}$, $\mathrm{SD}(y_{\cdot j''}) \mid
\mu_{j''},\tau_w,x_{j''})$; see the supplementary material, Section A.1
[\citet{ranta2015}]. The priors were the same as before, complemented
by the flat prior $\pi(\alpha)=\mathrm{U}(0,10^4)$. The marginal
posterior distribution of $(\sigma_b,\sigma_w)$ is shown at the right
in Figure~\ref{sigfig}. Compared with ``1/batch'' data, the variance
components are now better identified. However, to fully use the two
data sets jointly, a Bayesian evidence synthesis is applied below.

%s2.3.3 #&#
\subsubsection{Modeling based on combined data and results}

The posterior distribution was constructed by combining the likelihood
functions from the two data sets while keeping the same priors.
However, the common parameters made it possible to include ``1/batch''
data without the restrictive assumption of 100\% within-batch
prevalence. The corresponding factor of the likelihood was then written
as Binomial$(J' | N',q \alpha/(\alpha+2))$, since $E(p_{j'} \mid\alpha
) = \alpha/(\alpha+2)$.
The marginal posterior distribution for $(\sigma_b,\sigma_w)$ based on
the combined data is shown in the middle panel of Figure~\ref{sigfig}.
Both data sets were crucial for estimating the full set of parameters
in Table~\ref{restable}.

%s2.3.4 #&#
\subsubsection{Sensitivity to priors}

With ``$N_{j''}/\mathrm{batch}^+$'' data, the results are robust for the
reasonably uninformative prior choices because the data are informative
enough for the parameters. With ``1/batch'' data, the only sensitive
choice is the prior for the variance components because the data are
informative for the total variance. Uniform priors for the standard
deviations $\sigma_b$ and $\sigma_w$ were tested, and they led to quite
similar overall conclusions and point estimates. However, the
bimodality of marginal distributions was more pronounced with the Gamma
priors. If these data were the only data, then the uniform priors could
be preferred for robustness. In the evidence synthesis of the two data
sets, the choice of priors is less critical because the combined data
are fairly informative for all parameters. Because all of the results
and predictions were ultimately based on the evidence synthesis, the
default priors above were considered sufficient. With more seriously
limited data, the priors could have more effect. Simple uninformative,
or improper, standard priors might not work as such then, a known
pitfall for hierarchical models. The choice of prior becomes critical
for the between-batch variance $\sigma_b^2$ if the number of batches is
small and/or $\sigma_b^2$ is nearly zero [\citet{gelmanprior}].

%t2 #&#
\begin{table}
\tabcolsep=0pt
\caption{Summary of the parameter estimates from evidence synthesis
based on two data sets}
\label{restable}
\begin{tabular*}{\textwidth}{@{\extracolsep{4in minus 4in}}ld{2.2}d{2.4}d{3.2}@{}}
\hline
\multicolumn{1}{@{}l}{\textbf{Parameter}} & \multicolumn{1}{c}{\textbf{Mean}} & \multicolumn{2}{c@{}}{\textbf{95\% credible  interval}} \\
\hline
\multicolumn{4}{@{}l}{``1/batch'' data:} \\
$\mu$ & 2.1 & 1.9 & 2.3 \\
$q$ & 0.14 & 0.12 & 0.17 \\
$\sigma_b$ & 0.58 & 0.046 & 0.96 \\
$\sigma_w$ & 0.48 & 0.034 & 0.94 \\
$\phi=\sigma_w^2/(\sigma_w^2+\sigma_b^2)$ & 0.44 & 0.0015 & 1.0 \\[3pt]
\multicolumn{4}{@{}l}{Combined data:} \\
$\mu$ & 2.4 & 2.2 & 2.5 \\
$q$ & 0.15 & 0.12 & 0.18 \\
$\alpha$ & 85 & 38 & 177 \\
$\sigma_b$ & 0.66 & 0.52 & 0.82 \\
$\sigma_w$ & 0.74 & 0.68 & 0.80 \\[3pt]
$\phi=\sigma_w^2/(\sigma_w^2+\sigma_b^2)$ & 0.55 & 0.43 & 0.68 \\
\multicolumn{4}{@{}l}{``$N_{j''}/\mathrm{batch}^+$'' data:} \\
$\mu$ & 2.9 & 2.6 & 3.2 \\
$\alpha$ & 85 & 39 & 172 \\
$\sigma_b$ & 0.60 & 0.42 & 0.86 \\
$\sigma_w$ & 0.74 & 0.68 & 0.81 \\
$\phi=\sigma_w^2/(\sigma_w^2+\sigma_b^2)$ & 0.61 & 0.42 & 0.77\\
\hline
\end{tabular*}
\end{table}

%s2.4 #&#
\subsection{Consumer risk and microbiological criteria (MC)}
%s2.4.1 #&#
\subsubsection{Batch-specific inference, given the MC status of the batch}
Next we focus on predicting risk resulting from servings from a generic
new batch $j$.\vadjust{\goodbreak} The predictions depend on batch-specific parameters: the
batch contamination status $I_j$ (binary, ``$0/1$''), within-batch
prevalence $p_j$, and batch mean $\log_{10}$ concentration $\mu_j$ in
the contaminated carcasses.
These parameters have conditional distributions, given the previous parameters:
$I_j \mid q \sim \operatorname{Bern}(q)$ and $p_j \mid\alpha\sim
\operatorname{Beta}(\alpha,2)$ and $\mu_j \mid\mu,\tau_b \sim\mathrm{N}(\mu,\tau_b)$.
Therefore, for servings from a given batch the disease probability is
$P(\mathrm{ill}\mid\theta_j,\sigma_w)= I_j p_j P_0(\mathrm{ill} \mid\mu
_j,\sigma_w)$, conditional on batch parameters $\theta_j=(I_j,p_j,\mu
_j)$ and $\sigma_w$, with $P_0$ computed from the given QMRA model (see
the supplementary material, Section A.2 [\citet{ranta2015}]).

Alongside the risk associated with a batch, as expressed by the
batch-level parameters, we are also interested in the probability that
the batch complies with a particular microbiological criterion [\citet
{nauta}]. We have interpreted a stated criterion, defined by the
triplet $n/c/m$, as a condition for
accepting a batch. Hence, only batches where at most $c$ out of $n$
sampled carcasses exceed the contamination level of $m$ cfu/g are
accepted for consumption.

By taking the Bayesian approach, we treat the MC status as an
observation that provides additional evidence for a batch. This
knowledge subsequently updates the posterior distribution of the
parameters concerning such a batch. The posterior distribution of the
risk is then computed conditionally based on one of the following: (1)
the criterion is met (batch accepted), (2) the criterion is not met
(batch rejected), and (3) the criterion is not applied or,
equivalently, criterion status is not known. Based on the two data
sets, batches are accepted with a high probability: $P(\mathrm{MC\ is\ met}
\mid\mathrm{data1,\ data2})=0.95$. Knowing that the batch was accepted
leads to lower risk estimates for that batch. Without knowing the MC
status, the batch is contaminated with probability $P(I_j=1 \mid\mathrm
{data1,\ data2})=15\%$, but when the batch status is given, the
probability becomes either $P(I_j=1 \mid\mathrm{data1,\ data2,\ MC\
met})=10\%$ or $P(I_j=1 \mid\mathrm{data1,\ data2,\ MC\ not\ met})=1$.

Also, the observed batch status affects the probability of
concentrations on contaminated carcasses in the batch: $E( \mu_j \mid
\mathrm{data1,\ data2})=2.37$, but $E(\mu_j \mid\mathrm{data1,}$ $\mathrm{data2,\ MC\
not\ met})=2.96$. For the case ``MC not met,'' the batch is contaminated
with certainty (no false positives allowed). When the MC status is
unknown or the MC is met, it is possible that the batch is completely
free of contaminated carcasses. For the case ``MC met,'' it may be of
interest to compute the posterior mean of the mean concentration $\mu
_j$ depending also on this hidden variable: $E( \mu_j \mid\mathrm{data1,\
data2,\ MC\ met},I_j=1)=2.05$. Hence, for a compliant \textit{but
contaminated} batch, the mean concentration of contaminated carcasses
is probably lower than for a similar batch with unknown MC status.

%f4 #&#
\begin{figure}

\includegraphics{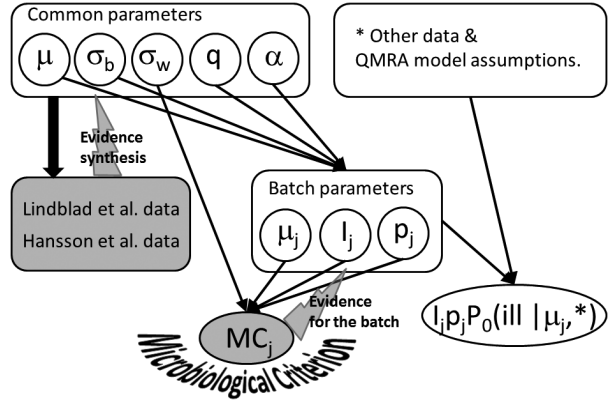}

\caption{Directed acyclic graph of the Bayesian model combining the two
carcass sample
data sets for common parameters and predicting the batch parameters
conditionally, now also
based on the status of the batch ($\mathrm{MC}_j=$ accepted/rejected).
The batch-specific illness probability (via QMRA model) then depends
both on the carcass sample data and
on the batch-specific status, for a generic new batch $j$ to be
predicted.} \label{bigdag}
\end{figure}

The posterior distribution of the batch parameters $\theta
_j=(I_j,p_j,\mu_j)$, conditional on given underlying parameters $q,\mu
,\sigma_w,\sigma_b,\alpha$ and the batch status ``MC met'' is $\pi
(\theta_j \mid\mathrm{MC\ met}, q,\mu,\sigma_w,\sigma_b,\alpha)$. By
Bayes's theorem, it is proportional to
%
%e3 #&#
\begin{equation}
P(\mathrm{MC\ met} \mid I_j,p_j,\mu_j,
\sigma_w) \operatorname{Bern}(I_j \mid q)\operatorname{Beta}(p_j
\mid\alpha,2)\mathrm{N}\bigl(\mu_j \mid\mu,\sigma_b^{-2}
\bigr).
\end{equation}
Because the underlying parameters are unknown, and because we had the
evidence from the two data sets, the marginal posterior distribution of
the batch parameters is
$\pi(\theta_j \mid\mathrm{MC\ met}, \mathrm{data1,\ data2} )$, which is
proportional to
%
%e4 #&#
\begin{eqnarray}
&&\int P(\mathrm{MC\ met} \mid I_j,p_j,\mu_j,
\sigma_w) \operatorname{Bern}(I_j \mid q) \operatorname{Beta}(p_j
\mid\alpha,2)\mathrm{N}\bigl(\mu_j \mid\mu,\sigma_b^{-2}
\bigr)
\nonumber
\\[-8pt]
\\[-8pt]
\nonumber
&&\qquad{}\times\pi(q,\mu,\sigma_w,\sigma_b,\alpha\mid\mathrm{data1,\
data2}) \,d (q,\mu,\sigma_w,\sigma_b,\alpha),
\end{eqnarray}
where the integral is taken over the underlying set of parameters
$(q,\mu,\sigma_w,\sigma_b$, $\alpha)$, representing, for example, a
country. Because observing MC status changes the probability for the
hidden batch status $I_j$ and
the log-cfu distribution, the added knowledge can also be expected to
affect the batch-specific consumer risk estimates;
see Figure~\ref{bigdag}. The posterior risk, given the MC status and
the data, is a single number,
%
%e5 #&#
\begin{eqnarray}
&&P(\mathrm{ill} \mid\mathrm{MC\ met},\mathrm{data1,\ data2} )
\nonumber
\\[-8pt]
\\[-8pt]
\nonumber
&&\qquad= \int_{\Theta_j}P(
\mathrm{ill} \mid\theta_j) \pi(\theta_j \mid\mathrm{MC\ met},
\mathrm{data1,\ data2} )\,d\theta_j,
\end{eqnarray}
resulting from integrating over all of the uncertain parameters, both
aleatoric and epistemic.

%t3 #&#
\begin{table}
\caption{Posterior means of $\mathit{RR}(q,\mu,\sigma_w,\sigma_b,\alpha)$, with
various MC ($n/c/m$). Default in bold.
$\mathit{RR}(q,\mu,\sigma_w,\sigma_b,\alpha)$ is risk conditional on acceptance
of the batch (MC is met)
divided by the risk conditional on ``MC not applied.'' Corresponding
means of rejection percentages are shown in subscripts. The same
results were obtained for $\mathit{RPR}$}
\label{RRs}
\begin{tabular*}{\textwidth}{@{\extracolsep{\fill}}lcccc@{}}
\hline
& \multicolumn{2}{c}{$\bolds{m=1000}$} & \multicolumn{2}{c@{}}{$\bolds{m=100}$} \\[-6pt]
& \multicolumn{2}{c}{\hrulefill} & \multicolumn{2}{c@{}}{\hrulefill} \\
$ $& $\bolds{n=5}$ & $\bolds{n=10}$ & $\bolds{n=5}$ & $\bolds{n=10}$ \\
\hline
$c=0$ & $0.20_{9\%}$ & $0.10_{11\%}$ & $0.01_{14\%}$ & $0.00_{14\%}$\\
${\mathbf{ c=1}}$ & ${\mathbf{0.42_{5\%}}}$ & $0.22_{8\%}$ & $0.06_{12\%}$ &
$0.01_{14\%}$\\
$c=2$ & $0.61_{3\%}$ & $0.34_{6\%}$ & $0.14_{10\%}$ & $0.03_{13\%}$ \\
$c=3$ & $0.78_{1\%}$ & $0.45_{5\%}$ & $0.30_{7\%}$ & $0.05_{12\%}$ \\
$c=4$ & $0.92_{0\%}$ & $0.56_{3\%}$ & $0.58_{4\%}$ & $0.09_{11\%}$\\
\hline
\end{tabular*}
\end{table}

%

%s3 #&#
\section{Comparisons of microbiological criteria using measures of
relative risk}

%s3.1 #&#
\subsection{Measures of relative risk}
Microbiological criteria were further studied by changing the values
for $n$, $c$ and $m$ in the criterion, as shown in Table~\ref{RRs}. In
comparisons, it is of interest to know the risk level of the accepted
batches relative to a situation where no criterion would be applied.
For this purpose, we defined the relative posterior risk ($\mathit{RPR}$) as the
ratio of the two batch-specific posterior probabilities:
%
%e6 #&#
\begin{equation}
\mathit{RPR} = \frac{P(\mathrm{ill} \mid\mathrm{MC\ met})}{P(\mathrm{ill} \mid\mathrm
{MC\ not\ applied})}.
\end{equation}
The two probabilities to be compared are single numbers obtained by
integrating over the parameter uncertainties. Alternatively, one could
study a parametric expression for relative risk,
%
%e7 #&#
\begin{equation}
\mathit{RR}(q,\mu,\sigma_w,\sigma_b,\alpha) =
\frac{P(\mathrm{ill} \mid\mathrm{MC\ met}, q,\mu,\sigma_w,\sigma_b,\alpha)}{
P(\mathrm{ill} \mid q,\mu,\sigma_w,\sigma_b,\alpha)}, \label{RR}
\end{equation}
as a function of the underlying (country-level) parameters $q,\mu,\sigma
_w,\sigma_b,\alpha$. To account for uncertain parameters, we would
compute the posterior mean
$E(\mathit{RR}(q,\mu, \sigma_w,\sigma_b,\alpha) \mid\mathrm{data}) $.
In practice, this computation requires 2D Monte Carlo: when the
underlying parameters $q,\mu,\sigma_w,\sigma_b,\alpha$ are sampled from
their posterior distribution using MCMC methods, within each iteration
step the batch parameters are Monte-Carlo integrated depending on the
current values of the underlying parameters.

The parametric approach to relative risk is much more
computer-intensive than $\mathit{RPR}$.
\citet{nauta} introduced a related measure, minimum relative residual
risk $\mathit{MRRR}$. In implementing it,
they assumed 100\% within-batch prevalence ($p=1$ for all batches) for
contaminated batches.
In our notation, we obtain the expression
%
%e8 #&#
\begin{eqnarray}
&&\mathit{MRRR}(q,\mu,\sigma_w,\sigma_b,p)\nonumber\\
 &&\qquad= \frac{q \int_{-\infty}^{\infty}
pP_0(\mathrm{ill} \mid\mu_j,\sigma_w)P(\mathrm{MC\ met} \mid p,\mu_j,\sigma
_w) \pi(\mu_j \mid\mu,\sigma_b) \,d\mu_j}{
P(\mathrm{ill} \mid q,\mu,\sigma_w,\sigma_b,p)}
\nonumber
\\[-8pt]
\\[-8pt]
\nonumber
&&\qquad= \frac{\int_{-\infty}^{\infty}
P(\mathrm{ill},\mathrm{MC\ met} \mid q,\mu_j,\sigma_w,p)\pi(\mu_j \mid\mu
,\sigma_b) \,d\mu_j}{
P(\mathrm{ill} \mid q,\mu,\sigma_w,\sigma_b,p)} \\
&&\qquad= \frac{P(\mathrm{ill},
\mathrm{MC\ met} \mid q,\mu,\sigma_w,\sigma_b,p)}{
P(\mathrm{ill} \mid q,\mu,\sigma_w,\sigma_b,p)},\nonumber
\end{eqnarray}
where $P(\mathrm{ill},\mathrm{MC\ met} \mid q,\mu,\sigma_w,\sigma_b,p)$ is,
in fact, the following total probability of illness:
%
%e9 #&#
\begin{eqnarray}
&&P(\mathrm{ill} \mid q,\mu,\sigma_w,\sigma_b,p,
\mathrm{intervention} )\nonumber
\\
&&\qquad= \underbrace{P(\mathrm{ill} \mid\mathrm{MC\ met}, q,\mu,\sigma_w,\sigma
_b,p)P(\mathrm{MC\ met} \mid q,\mu,\sigma_w,
\sigma_b,p)}_{ =P(\mathrm{ill}, \mathrm{MC\ met}\mid q,\mu,\sigma_w,\sigma_b,p)}
\nonumber
\\[-8pt]
\\[-8pt]
\nonumber
&&\qquad\quad{}+\underbrace{P(\mathrm{ill} \mid\mathrm{MC\ not\ met}, q,\mu,\sigma_w,
\sigma _b,p)}_{:=0, \mathrm{due\ to\ intervention}}\\
&&\qquad\quad{}\times P(\mathrm{MC\ not\ met}
\mid q,\mu,
\sigma_w,\sigma_b,p).\nonumber
\end{eqnarray}
This calculation assumes that contaminated batches are used, but only
after treatment that eliminates contamination.
$\mathit{MRRR}$ evaluates the quotient between the \emph{total probability} of
illness with such intervention and the total probability of illness
without an intervention.

As a point of comparison,
$\mathit{RR}$ evaluates the \emph{conditional probability} of illness for batches
where MC was met, divided by the probability of illness for batches
where MC was not applied. Therefore, if the same underlying parameter
values (``$\cdot$'') are used to evaluate the expressions, $\mathit{MRRR}(\cdot)
= \mathit{RR}(\cdot) \times P(\mathrm{MC\ met} \mid\cdot)$, so $\mathit{MRRR} \leq \mathit{RR}$. In
our example, these are nearly equal because $P(\mathrm{MC\ met} \mid\cdot
)\approx1$ with the Swedish data.
In earlier implementations of $\mathit{MRRR}$, the parameters $p,q,\mu,\sigma
_w,\sigma_b$ either are fixed values (e.g., $p=1$) or else result from
the assigned independent uncertainty distributions for each parameter,
but with $\mathit{RR}$ and $\mathit{RPR}$, the parameters are drawn from their joint
posterior distribution.

%s3.2 #&#
\subsection{Evaluating relative risks based on the posterior distribution}
To calculate the probability $P(\mathrm{ill} \mid q,\mu,\sigma_w,\sigma
_b,\alpha)$
in the denominator in equation (\ref{RR}) for $\mathit{RR}$, we can use the
following integral:
%
%e10 #&#
\begin{equation}
\int_{\Theta_j} P(\mathrm{ill} \mid\theta_j) \pi(
\theta_j \mid q,\mu ,\sigma_w,\sigma_b,
\alpha) \,d\theta_j,
\end{equation}
because illness is conditionally independent of $q,\mu,\sigma_w,\sigma
_b,\alpha$, given the batch parameters $\theta_j=(I_j,p_j,\mu_j)$.
The illness probability involves integrating the serving-specific
parameters $\theta_s$ [which include
number of bacteria from the broiler $y_c,n_c$, serving size $w$,
cross-contamination (transfer) probability $r$ in the salad making, and
dose $d$]. The whole expression can be approximated (see the
supplementary material, Section A.3 [\citet{ranta2015}]) as
%
%e11 #&#
\begin{equation}
\approx q \frac{\alpha}{\alpha+2} \frac{1}{L} \sum_{l=1}^{L}
\frac
{1}{M} \sum_{m=1}^{M}
P_0\bigl(\mathrm{ill} \mid\theta_s^{(m,l)}\bigr),
\end{equation}
where $\theta_s^{(m,l)}$ are Monte Carlo draws for the serving-specific
parameters within batches, sampled with the current values of $q,\mu
,\sigma_w,\sigma_b,\alpha$ at each MCMC iteration step, so that $\theta
_j^{(l)}$ is sampled first, then $\theta_s^{(m,l)}$ depending on each
$\theta_j^{(l)}$. $L$ batches and $M$ servings within each of the $L$
batches are simulated.

Next, to calculate the probability, $P(\mathrm{ill} \mid\mathrm{MC\ met},
q,\mu,\sigma_w,\sigma_b,\alpha)$, in the numerator in equation (\ref
{RR}) for the $\mathit{RR}$, we can write it as
%
%e12 #&#
\begin{equation}
\label{nume} \frac{P( \mathrm{ill}, \mathrm{MC\ met} \mid q,\mu,\sigma_w,\sigma_b,\alpha)}{
P(\mathrm{MC\ met} \mid q,\mu,\sigma_w,\sigma_b,\alpha)}.
\end{equation}
The denominator in equation (\ref{nume}), $P(\mathrm{MC\ met} \mid q,\mu
,\sigma_w,\sigma_b,\alpha)$, can be approximated (see the supplementary
material, Section A.3 [\citet{ranta2015}]) as
%
%e13 #&#
\begin{equation}
\approx q \frac{1}{L} \sum_{l=1}^L
1_{\{\mathrm{MC\ met}\}} \bigl(p_j^{(l)},\mu_j^{(l)},
\sigma_w\bigr) + (1-q),
\end{equation}
where the batch parameters are $L$ Monte Carlo draws from $\pi(\theta_j
\mid q,\mu,\sigma_w,\sigma_b$, $\alpha)$ and $1_{\{\mathrm{MC\
met}\}}$ is the indicator variable\vadjust{\goodbreak} for whether the batch complies, so
the average for the Monte Carlo sample is an approximation of the probability.
The numerator in equation (\ref{nume}), $P(\mathrm{ill},\mathrm{MC\ met}
\mid q,\mu,\sigma_w,\sigma_b,\alpha)$, can be approximated (see the
supplementary material, Section A.3 [\citet{ranta2015}]) as
%
%e14 #&#
\begin{equation}
\approx q \frac{1}{L} \sum_{l=1}^{L}
p_j^{(l)} \Biggl[ \frac{1}{M} \sum
_{m=1}^{M} P_0\bigl(\mathrm{ill} \mid
\theta_s^{(m,l)} \bigr) \Biggr] 1_{\{\mathrm{MC\ met}\}}
\bigl(p_j^{(l)},\mu_j^{(l)},
\sigma_w\bigr).
\end{equation}
With all of these Monte Carlo integrations, sampling batch parameters
and serving parameters within batches, we compute the approximation of
$\mathit{RR}(q,\mu,\sigma_w,\sigma_b,\alpha)$ within each step of the MCMC
simulation, which, in turn, draws samples for the underlying parameters
$(q,\mu,\sigma_w,\sigma_b,\alpha)$ from their posterior distribution.

%f5 #&#
\begin{figure}

\includegraphics{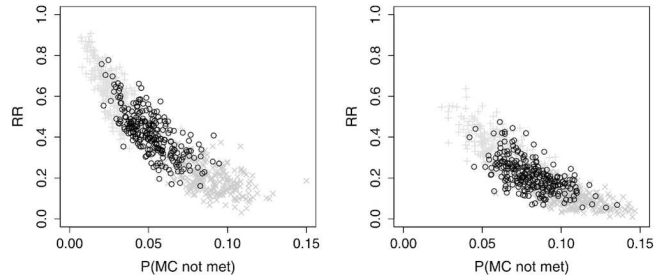}

\caption{Thinned samples from the joint posterior distributions of
$(P(\mathrm{MC\ not\ met}),\mathit{RR})$, both of which depend on the same unknown
parameters,
$(q,\mu,\sigma_w,\sigma_b,\alpha)$, for which the posterior
distributions were computed.
A single dot represents one MCMC draw of $q,\mu,\sigma_w,\sigma_b,\alpha
$ used in evaluating $P(\mathrm{MC\ not\ met})$ and $\mathit{RR}$.
Three MC with $m=1000$~cfu. Left: $\mathbf{n=5}$, $c=0$ (``$\times$''), $c=1$
(``$\circ$''), $c=2$ (``$+$'').
Right: $\mathbf{n=10}$, $c=0$ (``$\times$''), $c=1$ (``$\circ$''), $c=2$
(``$+$'').
Posterior means were $(0.09,0.20)$, $(0.05,0.42)$, $(0.03,0.61)$ and
$(0.11,0.10)$, $(0.08,0.22)$, and $(0.06,0.34)$, respectively.
MCMC run with 16,000 iterations, with 40 batches and 10 servings per
batch at each iteration.}
\label{MCs012}
\end{figure}

It is more efficient to compute the ratio of two posterior
probabilities, $\mathit{RPR}= P(\mathrm{ill} \mid\mathrm{MC\ met})/P(\mathrm{ill}
 \mid
\mathrm{MC\ not\ applied})$, than the posterior distribution 
of $\mathit{RR}$ or
$\mathit{MRRR}$, which requires 2D Monte Carlo. However, an advantage of 2D
Monte Carlo is that we then obtain credible intervals, for example, of
$\mathit{RR}(q,\mu,\sigma_w,\sigma_b,\alpha)$, which describe the uncertainty in
the underlying parameters. With the default MC $(n=5, c=1, m=1000)$,
the posterior mean was $E( \mathit{RR}(q,\mu,\sigma_w,\sigma_b,\alpha)
 \mid\mathrm
{data1,} \mathrm{data2}) \approx0.42$ and $\mathit{RPR} \approx0.40$; see the
supplementary material, Section A.4 [\citet{ranta2015}]. In our example
application, $E(\mathit{MRRR} \mid\mathrm{data1,\ data2})$ was also quite similar
($\approx0.39$). Ideally, the percentage of rejected batches (with
``MC not met'') in the total production and the relative risk ($\mathit{RR}$)
should both be low. The 2D-uncertainty plot for these is shown in
Figure~\ref{MCs012}. The final result for a particular criterion, which
accounts for all uncertainties, can be obtained by taking the overall
posterior means.

%s4 #&#
\section{Discussion}

The uncertainties of the risk estimates emerge roughly for two
qualitatively different reasons:
(1) existing but partial or limited data from the specific production
system and (2) external assumptions.
The latter cannot easily be avoided when microbiological risk
assessments aim to cover production chains and processes ranging from
food production to consumption.
Here we focused on uncertainties that can be quantified more explicitly
based on production sample data. This analysis was illustrated with two
data sets that contained partial but complementary evidence.

The posterior distribution of the core parameters was used to predict
the consequent risk for consumers, so the uncertainties were propagated
into the final risk estimates. However, this assessment is contingent
upon the often unquantifiable uncertainties
concerning the QMRA model for the remaining food pathway. Our approach
can be combined with any available QMRA model.
Parallel to the posterior predictive consumer risk, we also predicted
the outcome of a batch-specific microbiological criterion, as defined
by the triplet $n/c/m$.
The parametric risk, which depends on the acceptance of batches, as
determined by the MC, divided by the parametric risk without any MC at
all, was described by the relative risk $\mathit{RR}(q,\mu,\sigma_w,\sigma
_b,\alpha)$. When comparing MC in Figure~\ref{MCs012}, it is evident
how much $\mathit{RR}$ can be reduced by increasing the sample size $n$ or by
decreasing the number of positives $c$ allowed to exceed the
concentration $m$ among them. At the same time, the expected percentage
of noncompliant batches increases, which will increase costs if
noncompliant batches are rejected or otherwise specially treated. It is
then possible to optimize the criterion to achieve a higher risk
reduction with tolerable costs. In this way, the chosen criterion would
also be risk-based [\citet{nauta}].
By comparison, some MC are nearly equivalent. For example, both ``$n=5,
c=2, m=1000$'' and ``$n=10, c=4, m=1000$'' have $\mathit{RR} \approx0.6$ and an
expected rejection percentage of $3\%$. Of course, the latter has a
higher sampling cost because of double the number of samples. Further,
$\mathit{RPR}$ and $E(\mathit{RR} \mid\mathrm{data1,\ data2})$ only describe relative
effects. If the absolute risk level is already low, statistically
significant reductions in relative risks might not be epidemiologically
significant. The burden of disease, in number of cases, is more
difficult to judge than $\mathit{RR}$'s, because of uncertainties along the
carcass-to-serving path; see \citet{ternhag}.

Computationally, $\mathit{RR}$ depends on the unknown parameters and involves 2D
Monte Carlo. The simulation of batches and of servings within batches
needs some optimization and can be slow to run. Instead, relative
posterior risk $\mathit{RPR}$ is much faster to compute and gives practically
the same result. The same posterior illness probability can be obtained
either by computing $E_d(P_0(\mathrm{ill} \mid d))$ by Monte Carlo at
each MCMC iteration and then taking the average over iterations or
simply by computing $P_0(\mathrm{ill} \mid d)$ with all the current\vadjust{\goodbreak}
parameters at each MCMC iteration and then averaging over iterations.
Eventually, \emph{all} unknown parameters will become integrated when
computing the final posterior probability. The order of integration
does not matter, and it is more efficient not to run the 2D Monte Carlo
within the MCMC. Code for the models with sample data is freely
available, so
that risk assessors can compute results that explicitly depend on their
own data.
Since the model code contains two qualitatively different examples,
with individual and summary data, it could easily accommodate typical
situations.

\section*{Acknowledgments}
We thank the Nordic Working Group for Microbiology~\& Animal Health and
Welfare (NMDD)
\surl{www.norden.org/en/}
for funding, and the editors for very careful and detailed comments
that greatly improved the manuscript.

\begin{supplement}[id=suppA]
\stitle{Appendix: A Bayesian approach to the evaluation of risk-based
microbiological criteria for \textit{Campylobacter} in broiler meat\\}
\slink[doi]{10.1214/15-AOAS845SUPP} %[doi,text={...}] - jei reikia
%suskaldyti doi
\sdatatype{.pdf}
\sfilename{aoas845\_supp.pdf}
\sdescription{More details of computations and the BUGS codes are
described in the supplementary materials.}
\end{supplement}

% imsref loaded by akundreckaite, 2015-07-14 16:17:46
% imsref loaded by akundreckaite, 2015-07-15 08:02:31
%

%\begin{appendix}
%\section{}
%\end{appendix}

%\section*{Acknowledgments}

%\begin{thebibliography}{99}
%\bibitem[\protect\citeauthoryear{}{()}]{r1}
%\bibitem{r1}
%\end{thebibliography}

\printaddresses
\end{document}